\documentclass[aip,jmp,preprint,onecolumn,superscriptaddress]{revtex4-1}
\usepackage{amsfonts}
\usepackage{amssymb}
\usepackage{graphicx}
\usepackage{amsmath}
\usepackage{mathrsfs}
\usepackage{amsthm}
\usepackage{bbm}
\usepackage{hyperref}
\usepackage{makeidx}
\usepackage{epsfig}
\usepackage[cp1251]{inputenc}

\newtheorem{theorem}{Theorem}

\newtheorem{corollary}[theorem]{Corollary}

\newtheorem{lemma}[theorem]{Lemma}
\newtheorem{proposition}[theorem]{Proposition}

\begin{document}

\title{Accessible information of a general quantum Gaussian ensemble}
\author{A. S. Holevo}
\affiliation{Steklov Mathematical Institute, 119991 Moscow, Russia.}
\email{holevo@mi-ras.ru}
\date{}

\begin{abstract}
Accessible information, which is a basic quantity in quantum information
theory, is computed for a general quantum Gaussian ensemble under certain
``threshold condition''. It is shown that the maximizing measurement is
Gaussian, constituting a far-reaching generalization of the optical
heterodyning. This substantially extends the previous result
concerning the gauge-invariant case, even for a single bosonic mode. A
simple sufficient condition is provided that implies the threshold condition
for general Gaussian ensemble. The results are illustrated on the
single-mode case.
\end{abstract}

\maketitle

\section{Introduction}

Accessible information of an ensemble of quantum states is a basic quantity
in quantum information theory: it is equal to the maximal amount of the
Shannon information which can be gained from a given quantum ensemble (a
collection of ``signal'' quantum states with fixed probabilities) in a
one-step measurement. This quantity is often difficult to compute, the
problem lies in finding the \textit{global maximum } of a \textit{convex}
functional, when the maximizer turns out to be highly non-unique and the
standard tools of convex analysis become inefficient. The problem becomes
still more complicated for continuous variable (CV) systems which constitute
one of the prospective platforms for implementation of ideas of quantum
information theory (see e.g. \cite{sera}). The quantum Shannon theory for CV
systems requires mathematical tools of infinite-dimensional Hilbert spaces
and symplectic vector spaces, see \cite{QSCI}.

The present paper is a continuation and extension of our paper \cite{acc}
which gave a solution for the problem going back to 1970-s: it was shown
there that accessible information of a gauge-invariant bosonic Gaussian
ensemble is attained by a multimode generalization of heterodyne
measurement, and hence can be computed exactly. (Loosely speaking, gauge
invariance means that the problem has a unique natural complex structure. In
quantum optics, this is related to phase-insensitivity of the system.)

In the present paper we extend this result to arbitrary Gaussian ensembles
satisfying certain \textquotedblleft threshold condition\textquotedblright\
. This condition is the one that allows to reduce the classical capacity
problem to a simpler minimum output entropy problem, and it is always
fulfilled in the particularly tractable gauge-invariant case. Thus we obtain
here a ``Gaussian maximizer'' result in a situation going beyond gauge
invariance (which is often assumed, see e.g.\ \cite{depalma}, \cite{QSCI}
for various aspects of the famous ``Gaussian optimizer conjecture'' in
analysis and quantum information theory). Main tools will be the
infinite-dimensional version of \textquotedblleft ensemble-observable
duality\textquotedblright\ developed in \cite{acc} and the multiplication
formulas for Gaussian operators from \cite{lami} (see also Appendix 1).


\section{Preliminaries}

We refer reader to \cite{QSCI} for definitions of basic notions of quantum
statistics. Let $\mathcal{H}$ be a separable Hilbert space, $\mathcal{X}$ a
standard measurable space. An \textit{ensemble} $\mathcal{E}=\left\{ \pi
(dx),\rho _{x}\right\} $ consists of a probability measure $\pi (dx)$ on $%
\mathcal{X}$ \ and a measurable family of density operators (quantum states)
$x\rightarrow \rho _{x}$ on $\mathcal{H}$. The \textit{average state} of the
ensemble is the barycenter of this measure%
\begin{equation*}
\bar{\rho}_{\mathcal{E}}=\int_{\mathcal{X}}\rho _{x}\,\pi (dx),
\end{equation*}%
the integral existing in the strong sense in the Banach space of trace-class
operators on $\mathcal{H}$. Let $M=\{M(dy)\}$ be an \textit{observable}
(probability operator-valued measure = POVM) on $\mathcal{H}$ with the
outcome space $\mathcal{Y}$ . There exists a $\sigma -$finite measure $\mu
(dy)$ such that for any density operator $\rho $ the probability measure $%
\mathrm{Tr}\rho M(dy)$ is absolutely continuous w.r.t. $\mu (dy),$ thus
having the probability density $p_{\rho }(y)$ (one can take $\mu (dy)=%
\mathrm{Tr}\rho _{0}M(dy)$ where $\rho _{0}$ is a nondegenerate density
operator).

The joint probability distribution of $x,y$ on $\mathcal{X\times Y}$ is
uniquely defined by the relation%
\begin{eqnarray*}
P(A\times B)&=&\int_{A}\pi (dx)\mathrm{Tr}\,\rho _{x}M(B) \\
&=&\mathrm{Tr}\int_{A}\int_{B}\,p_{\rho _{x}}(y)\,\pi (dx)\mu (dy),
\end{eqnarray*}%
where $A$ is an arbitrary Borel subset of $\mathcal{X}$ and $B$ is that of $%
\mathcal{Y}.$ The classical Shannon information between $x,y$ is equal to%
\begin{eqnarray*}
I(\mathcal{E},M) &=&\int \int \pi (dx)\mu (dy)p_{\rho _{x}}(y)\log \frac{%
p_{\rho _{x}}(y)}{p_{\bar{\rho}_{\mathcal{E}}}(y)} \\
&=&h(p_{\bar{\rho}_{\mathcal{E}}})-\int h(p_{\rho _{x}})\pi (dx),
\end{eqnarray*}%
where
\begin{equation*}
h(p)=-\int p(x)\log p(x)\mu (dx)
\end{equation*}%
is the differential entropy of a probability density $p(x).$ There is a
special class of probability densities we will be dealing with for which the
differential entropy is well-defined (see \cite{acc} for the detail).

The \textit{accessible information} of the ensemble $\mathcal{E}$ is defined
as%
\begin{equation}
A(\mathcal{E}\mathbf{)=}\sup_{M}I(\mathcal{E},M),  \label{ai}
\end{equation}%
where the supremum is over all observables $M$ on $\mathcal{H}$.

We will systematically use notations and results from the book \cite{QSCI}.
Consider the finite-dimensional symplectic space $(Z,\Delta )$ with $Z=%
\mathbb{R}^{2s}$ and
\begin{equation}
\Delta =\mathrm{diag}\left[
\begin{array}{cc}
0 & 1 \\
-1 & 0%
\end{array}%
\right] _{j=1,\dots ,s}.  \label{delta}
\end{equation}%
In what follows $\mathcal{H}$ will be the space of an irreducible
representation $z\rightarrow W(z);\,z\in Z,$ of the \textit{canonical
commutation relations} 
\begin{equation}  \label{weyl}
W(z)W(z^{\prime })=\exp [-\frac{i}{2}z^{t}\Delta z^{\prime }]\,W(z+z^{\prime
}).
\end{equation}
Here $W(z)=\exp i\,Rz$ are the unitary Weyl operators with the generators
\begin{equation}
Rz=\sum_{j=1}^{s}(x_{j}q_{j}+y_{j}p_{j}),  \label{cano}
\end{equation}%
$z=[x_{j}\,\,y_{j}]_{j=1,\dots ,s}^{t}$, and $R=[q_{j}\,\,p_{j}]_{j=1,\dots
,s}$ are the canonical observables of the quantum system in question
satisfying $q_{j}p_{k}-p_{k}q_{j}=\delta _{jk}I$ . In quantum communication
theory they describe the relevant modes of the field on receiver's aperture
(see, e.g. \cite{sera}). The \textit{displacement operators} $D(z)=W(-\Delta
^{-1}z)$ satisfy the equation that follows from the canonical commutation
relations (\ref{weyl})
\begin{equation}
D(z)^{\ast }W(w)D(z)=\exp \left( iw^{t}z\right) W(w).  \label{dis}
\end{equation}

A centered \textit{Gaussian state} $\rho _{\alpha }$ is determined by its
quantum characteristic function%
\begin{equation}  \label{gs}
\mathrm{Tr}\,\rho _{\alpha }W(z)=\exp \left( -\frac{1}{2}z^{t}\alpha
z\right) ,
\end{equation}%
where the \textit{covariance matrix} $\alpha =\mathrm{Re\,}\mathrm{Tr}%
\,R^{t}\rho R$ is a real symmetric $2s\times 2s$-matrix satisfying
\begin{equation}
\alpha \geq \pm \frac{i}{2}\Delta .  \label{ur}
\end{equation}

Operator $J$ in $(Z,\Delta )$ is called \textit{operator of complex
structure } if
\begin{equation}
J^{2}=-I_{2s},  \label{j2e}
\end{equation}%
where $I_{2s}$ is the identity operator in $Z$, and it is $\Delta -$positive
in the sense that
\begin{equation}
\Delta J=-J^{t}\Delta ,\quad \Delta J\geq 0.  \label{comstr}
\end{equation}
In other words, $\Delta $ is \textit{tamed} by $J$.

The Gaussian state $\rho _{\alpha }$ is pure if and only if $\alpha =\frac{1%
}{2}\Delta J$ where $J$ is an operator of complex structure. Such state is
called $J-$vacuum and denoted $\rho _{\frac{1}{2}\Delta J}.$ The
non-centered pure states $D(z)\rho _{\frac{1}{2}\Delta J}D(z)^{\ast }$ are
called $J-$coherent states (see sec. 12.3.2 of ~\cite{QSCI}).

Consider the operator $A=\Delta ^{-1}\alpha $. The operator $A$ is
skew-symmetric in the Euclidean space $(Z,\alpha )$ with the scalar product $%
\alpha (z,z^{\prime })=z^{t}\alpha z^{\prime }$. According to a theorem from
linear algebra, there is an orthogonal basis $\left\{ e_{j},h_{j}\right\} $
in $(Z,\alpha )$ and positive numbers $\left\{ \alpha _{j}\right\} $ (called
symplectic eigenvalues of $\alpha$) such that
\begin{equation*}
Ae_{j}=\alpha _{j}h_{j};\quad Ah_{j}=-\alpha _{j}e_{j}, \quad j=1,\dots,s.
\end{equation*}%
Inequality (\ref{ur}) is equivalent to $N_{j}\equiv \alpha _{j}-1/2\geq 0$, $%
j=1,\dots,s.$ Choosing the normalization~ $\alpha (e_{j},e_{j})=\alpha
(h_{j},h_{j})=\alpha _{j}$ gives a symplectic basis in $(Z,\Delta )$.

There is an operator of complex structure, commuting with the operator $%
A=\Delta ^{-1}\alpha ,$ namely, the orthogonal operator $J_{\alpha }$ from
the polar decomposition
\begin{equation}
A=\left\vert A\right\vert J_{\alpha }=J_{\alpha }\left\vert A\right\vert
\label{kspd}
\end{equation}%
in the Euclidean space $(Z,\alpha ).$ The action of $\left\vert A\right\vert
$ and $J_{\alpha }$ in the symplectic basis $\left\{ e_{j},h_{j}\right\} $
constructed above is given by the formula
\begin{eqnarray*}
\left\vert A\right\vert e_{j} &=&\alpha _{j}e_{j},\quad \left\vert
A\right\vert h_{j}=\alpha _{j}h_{j}; \\
J_{\alpha }e_{j} &=&h_{j},\quad \quad J_{\alpha }h_{j}=-e_{j}.
\end{eqnarray*}%
Inequality (\ref{ur}) is equivalent to
\begin{equation}
\alpha \geq \frac{1}{2}\Delta J_{\alpha }  \label{ur1}
\end{equation}
because it amounts to $\alpha _{j}-1/2\geq 0$, $j=1,\dots,s.$

We will consider the general Gaussian observable (probability
operator-valued measure = POVM) on $Z=\mathbb{R}^{2s} $ (see ~\cite{acc})
\begin{equation}
\widetilde{M}(d^{2s}z)=D(Kz)\rho _{\beta }D(Kz)^{\ast }\frac{\left\vert \det
K\right\vert \,d^{2s}z}{\left( 2\pi \right) ^{s}},  \label{MTA}
\end{equation}
where $K$ is a nondegenerate real matrix and $\rho _{\beta }$ is a centered
Gaussian density operator with the real symmetric covariance matrix $\beta .$
In this case $\mu $ is just the normalized Lebesgue measure on $Z=\mathbb{R}%
^{2s}.$ Especially important is the case $K=I$ where
\begin{equation}
M(d^{2s}z)=D(z)\rho _{\beta }D(z)^{\ast }\frac{d^{2s}z}{\left( 2\pi \right)
^{s}}.  \label{MTB}
\end{equation}
The probability density of the observable (\ref{MTB}) in the state $\rho
_{\alpha }$ is computed by using the Parceval formula for the quantum
Fourier transform (see ~\cite{acc-noJ})
\begin{eqnarray}
p_{\rho _{\alpha }}(z) &=&\mathrm{Tr}\,\rho _{\alpha }D(z)\rho _{\beta
}D(z)^{\ast }  = \int \exp \left( -\frac{1}{2}w^{t}\alpha w\right) \exp \left( -iw^{t}z-%
\frac{1}{2}w^{t}\beta w\right) \frac{\,d^{2s}w}{\left( 2\pi \right) ^{s}}
\notag \\
&=&\frac{1}{\sqrt{\left( 2\pi \right) ^{s}\mathrm{\det }\left( \alpha +\beta
\right) }}\exp \left( -\frac{1}{2}z^{t}\left( \alpha +\beta \right)
^{-1}z\right) .  \label{plus}
\end{eqnarray}
An important special case of observable (\ref{MTB}) is the (squeezed)
heterodyne measurement
\begin{equation}
M(d^{2s}z)=D(z)\rho _{\frac{1}{2}\Delta J_{\beta }}D(z)^{\ast }\frac{d^{2s}z%
}{\left( 2\pi \right) ^{s}}.  \label{vacM}
\end{equation}%
(see Appendix of ~\cite{acc} for the gauge-invariant case). Then (\ref{MTB})
can be considered as noisy version of the heterodyne measurement, and (\ref%
{MTA}) -- as (matrix) rescaling of (\ref{MTB}), which describes classical
linear post-processing of the measurement outcomes.

\section{The main result}

We first prove the lemma:

\begin{lemma}
\label{l1} \textit{Let $\widetilde{M}$ be the Gaussian observable (\ref{MTA}%
) where $\rho _{\beta }$ is a centered Gaussian density operator with the
real symmetric covariance matrix $\beta .$ Assume that $\alpha $ is
covariance matrix of a Gaussian state }$\rho _{\alpha }$\textit{\ satisfying
the condition }
\begin{equation}
{\alpha }\geq \frac{1}{2}\Delta J_{\beta }.  \label{thr}
\end{equation}%
\textit{Then} 
\begin{eqnarray}
\max_{\mathcal{E}:\bar{\rho}_{\mathcal{E}}=\rho _{\alpha }}I(%
\mathcal{E},\tilde{M})&=&\frac{1}{2}\log \mathrm{\det }\left( \alpha +\beta
\right)-\frac{1}{2}\log \mathrm{\det }\left( \beta +\frac{1}{2}\Delta
J_{\beta }\right)   \label{CXA} \\
&=&\frac{1}{2}\log \mathrm{\det }\left( \alpha +\beta \right) \left( \beta +%
\frac{1}{2}\Delta J_{\beta }\right) ^{-1},  \notag
\end{eqnarray}%
\textit{which is attained on the ensemble }$\mathcal{E}_{\ast }$ \textit{of $%
J_{\beta }-$coherent states $D(z)\rho _{\frac{1}{2}\Delta J_{\beta
}}D(z)^{\ast }$, where $z$ has the centered Gaussian probability
distribution }$\pi _{\gamma }$ \textit{with the covariance matrix}
\begin{equation}
\gamma ={\alpha }-\frac{1}{2}\Delta J_{\beta }.  \label{gamma}
\end{equation}
\end{lemma}

We would like to stress that in this paper we do not assume the gauge
symmetry: $\alpha $ and $\beta $ need not share the common complex
structure, $J_{\alpha }$ need not coincide with $J_{\beta }.$ In the
gauge-invariant case, where the complex structure is unique, we have the
correspondence~\cite{QSCI} $J_{\alpha }=J_{\beta }=\Delta ^{-1}\rightarrow i$%
, $\alpha \rightarrow \Sigma +I_{s}/2$, $\beta \rightarrow N+I_{s}/2,$ $%
\Delta ^{-1}\beta \rightarrow i\left( N+I_{s}/2\right) $, \ and (\ref{CXA})
turns into the formula of theorem 1 in ~\cite{acc}:
\begin{equation}
C_{\chi }(\widetilde{M};\Sigma )=\log \det \left( I_{s}+\left(
N+I_{s}\right) ^{-1}\Sigma \right) .  \label{cchig}
\end{equation}%
\textit{Proof (sketch).} We will need the formula for the differential
entropy of a multidimensional Gaussian probability density $p_{\gamma }$
with the covariance matrix $\gamma :$
\begin{equation}
h(p_{\gamma })=\frac{1}{2}\log \det \gamma +C,  \label{dent}
\end{equation}%
where the constant $C$ depends on the normalization of the Lebesgue measure
involved in the definition of the differential entropy (cf. ~\cite{cover}).

In ~\cite{acc} it is shown that the result does not depend on $K$ so that we
can take $K=I_{2s}$ and consider the POVM (\ref{MTB}). Then the proof is
parallel to proof of theorem 1 in ~\cite{acc-noJ}. We have
\begin{equation}
\max_{\mathcal{E}:\bar{\rho}_{\mathcal{E}}=\rho _{\alpha }}I(\mathcal{E},%
\tilde{M})=h\left( p_{\rho _{\alpha }}\right) -\min_{\rho }h\left( p_{\rho
}\right) .  \label{maxI}
\end{equation}%
Let us show that the maximum is attained on the ensemble
\begin{equation*}
\mathcal{E}_{\ast }=\left\{ {\pi }_{\gamma }(dz),D(z)\rho _{\frac{1}{2}%
\Delta J_{\beta }}D(z)^{\ast }\right\} ,
\end{equation*}%
with $\gamma $ given by (\ref{gamma}). The condition (\ref{thr}) ensures
existence of the centered Gaussian distribution ${\pi }_{\gamma }(dz)$ on $Z$
with the covariance matrix $\gamma ={\alpha }-\frac{1}{2}\Delta J_{\beta }.$
The average state is%
\begin{equation*}
\bar{\rho}_{\mathcal{E}}=\int_{\mathbb{R}^{2s}}D(z)\rho _{\frac{1}{2}\Delta
J_{\beta }}D(z)^{\ast }\,{\pi }_{\gamma }(dz)=\rho _{\alpha },
\end{equation*}%
One can check this equality by computing the quantum characteristic
functions. The probability density of (\ref{MTB}) is given by (\ref{plus}).
Thus according to (\ref{dent})
\begin{equation}
h\left( p_{\rho _{\alpha }}\right) =\frac{1}{2}\log \mathrm{\det }\left(
\alpha +\beta \right) +C.  \label{hps}
\end{equation}

The result of the paper ~\cite{ghm} (Proposition 4; see also ~\cite{acc})
concerning the minimal output entropy of the Gaussian measurement channel
implies that the minimizer can be taken as the vacuum state $\rho _{\frac{1}{%
2}\Delta J_{\beta }}$ related to the complex structure $J_{\beta }$.
Substituting $\alpha =\frac{1}{2}\Delta J_{\beta }$ into (\ref{hps}), we get
\begin{eqnarray}
\min_{\rho }h\left( p_{\rho }\right)& =&h\left( p_{\rho _{\frac{1}{2}\Delta
J_{\beta }}}\right)  \label{mine} \\
& =&\frac{1}{2}\log \mathrm{\det }\left( \beta +\frac{1}{2}\Delta J_{\beta
}\right) +C.  \notag
\end{eqnarray}%
Substituting (\ref{hps}) and (\ref{mine}) into (\ref{maxI}), we get (\ref%
{CXA}). $\square $

Now we can prove the main result of the paper.

\begin{theorem}
\label{t1} \textit{Let $\gamma $ be a real positive definite matrix and let $%
\mathcal{E}$ be the Gaussian ensemble $\left\{ \pi _{\gamma }(d^{2s}z),\rho
_{\beta ,z}\right\} ,$ where}
\begin{eqnarray}
\pi _{\gamma }(d^{2s}z) &=&\exp \left( -\frac{1}{2}z^{\ast }\gamma
^{-1}z\right) \frac{d^{2s}z}{\left( 2\pi \right) ^{s}\sqrt{\det \gamma }}%
,\quad  \label{GE} \\
\rho _{\beta ,z} &=&D(z)\rho _{\beta }D(z)^{\ast }  \label{GE2}
\end{eqnarray}

\textit{Then the accessible information (\ref{ai}) of this ensemble is equal
to
\begin{equation}
A(\mathcal{E}\mathbf{)}=\frac{1}{2}\log \mathrm{\det }\left( \tilde{\alpha}+%
\tilde{\beta}\right) \left( \tilde{\beta}+\frac{1}{2}\Delta J_{\tilde{\beta}%
}\right) ^{-1},  \label{aim}
\end{equation}%
where
\begin{eqnarray}
\tilde{\alpha} &=&\gamma +\beta ,  \label{tildea} \\
\tilde{\beta} &=&\tilde{\alpha}\tilde{\Upsilon}^{t}\gamma ^{-1}\tilde{%
\Upsilon}\tilde{\alpha}-\tilde{\alpha},  \label{tildeb} \\
\tilde{\Upsilon} &=&\sqrt{I_{2s}+\left( 2\tilde{\alpha}\Delta ^{-1}\right)
^{-2}},  \label{ips}
\end{eqnarray}%
provided the threshold condition
\begin{equation}
\tilde{\alpha}-\frac{1}{2}\Delta J_{\tilde{\beta}}\geq 0  \label{cond}
\end{equation}%
holds.}

\textit{The supremum in (\ref{ai}) is attained on the squeezed heterodyne
observable
\begin{equation}
M_{\ast }(d^{2s}z)=D(Kz)\rho _{\beta _{\ast }}D(Kz)^{\ast }\frac{\left\vert
\det K\right\vert d^{2s}z}{\left( 2\pi \right) ^{s}},  \label{oo}
\end{equation}%
where $K$ is a nondegenerate matrix and} 
\begin{equation}
\beta _{\ast }=\tilde{\alpha}\tilde{\Upsilon}^{t}\left( \tilde{\alpha}-\frac{%
1}{2}\Delta J_{\tilde{\beta}}\right) ^{-1}\tilde{\Upsilon}\tilde{\alpha}-%
\tilde{\alpha}.  \label{opt}
\end{equation}%
\end{theorem}

Notice that the condition (\ref{cond}) is automatically fulfilled in the
gauge-invariant case where the complex structure is unique: $J_{\tilde{\beta}%
}=J_{\alpha }=J_{\beta }=\Delta ^{-1}\rightarrow i$, and the statement
reduces to theorem 2 in ~\cite{acc}. Otherwise, apart from the single-mode
case considered in the following section, the condition (\ref{cond}) might
be difficult to check, therefore the following simple sufficient condition
could be useful.

\begin{proposition}
\label{p1} \textit{If $\gamma \geq \beta $ then (\ref{cond}) holds.}
\end{proposition}

\textit{Proof.} Consider the inequality%
\begin{equation*}
\tilde{\alpha}\geq \tilde{\Upsilon}\tilde{\alpha}\tilde{\Upsilon}^{t},
\end{equation*}%
which amounts to $I\geq A^{t}A,$ where
\begin{equation*}
A=\tilde{\alpha}^{1/2}\tilde{\Upsilon}^{t}\tilde{\alpha}^{-1/2}=\sqrt{I_{2s}-%
\frac{1}{4}B^{t}B}=A^{t},
\end{equation*}%
with $B=\tilde{\alpha}^{-1/2}\Delta \tilde{\alpha}^{-1/2}.$

Then the inequality $\gamma \geq \beta $ and (\ref{tildea}) imply
consecutively
\begin{equation*}
2\gamma \geq \tilde{\Upsilon}\tilde{\alpha}\tilde{\Upsilon}^{t},
\end{equation*}%
\begin{equation*}
\tilde{\Upsilon}^{t}\gamma ^{-1}\tilde{\Upsilon}\leq 2\tilde{\alpha}^{-1},
\end{equation*}%
\begin{equation*}
\tilde{\alpha}\geq \tilde{\alpha}\tilde{\Upsilon}^{t}\gamma ^{-1}\tilde{%
\Upsilon}\tilde{\alpha}-\tilde{\alpha}=\tilde{\beta}.
\end{equation*}
But $\tilde{\beta}\geq \frac{1}{2}\Delta J_{\tilde{\beta}}$ , which implies (%
\ref{cond}). $\square $

\textit{Proof of theorem \ref{t1}}. For the clarity of proofs we assume that
the covariance matrix $\gamma $ of the Gaussian distribution $\pi _{\gamma }$
is nondegenerate, although this restriction can be relaxed by using more
formal computations with characteristic functions. By using the
characteristic function and (\ref{dis}), we find the average state of the
ensemble $\mathcal{E}$
\begin{equation}
\bar{\rho}_{\mathcal{E}}\equiv \int \rho _{\beta ,z}\,\pi _{\gamma
}(d^{2s}z)=\rho _{\gamma +\beta }=\rho _{\tilde{\alpha}}.  \label{aver}
\end{equation}

Proof of (\ref{aim}) uses ensemble-observable duality from ~\cite{acc},
which is sketched below (see ~\cite{acc} for detail of mathematically
rigorous description).

Let $\mathcal{E}=\left\{ \pi (dx),\rho _{x}\right\} $ be an ensemble, $\mu
(dy)$ a $\sigma -$finite measure and $M=\left\{ M(dy)\right\} $ an
observable having operator density $m(y)=M(dy)/\mu (dy)$ with values in the
algebra of bounded operators in $\mathcal{H}$. The dual pair
ensemble-observable $(\mathcal{E}^{\prime },M^{\prime })$ is defined by the
relations
\begin{eqnarray}
\mathcal{E}^{\prime }:\quad \pi ^{\prime }(dy)&=&\mathop{\rm Tr}\nolimits%
\bar{\rho}_{\mathcal{E}}\,M(dy),  \notag \\
\rho _{y}^{\prime }&=&\frac{\bar{\rho}_{\mathcal{E}}^{1/2}m(y)\bar{\rho}_{%
\mathcal{E}}^{1/2}}{\mathop{\rm Tr}\nolimits\bar{\rho}_{\mathcal{E}}\,m(y)};
\label{piprime}
\end{eqnarray}
\begin{equation}
M^{\prime }:\quad M^{\prime }(dx)=\bar{\rho}_{\mathcal{E}}^{-1/2}\rho _{x}%
\bar{\rho}_{\mathcal{E}}^{-1/2}\pi (dx),  \label{mprime}
\end{equation}%
Then the average states of both ensembles coincide
\begin{equation}
\bar{\rho}_{\mathcal{E}}=\bar{\rho}_{\mathcal{E}^{\prime }}  \label{III}
\end{equation}%
and the joint distribution of $x,y$ is the same for both pairs $(\mathcal{E}%
,M)$ and $(\mathcal{E}^{\prime },M^{\prime })$ so that%
\begin{equation}
I(\mathcal{E},M)=I(\mathcal{E}^{\prime },M^{\prime }).  \label{II}
\end{equation}%
Moreover,
\begin{equation}
\sup_{M}I(\mathcal{E},M)=\sup_{\mathcal{E}^{\prime }:\bar{\rho}_{\mathcal{E}%
^{\prime }}=\bar{\rho}_{\mathcal{E}}}I(\mathcal{E}^{\prime },M^{\prime }),
\label{inf}
\end{equation}%
where the supremum in the right-hand side is taken over all ensembles $%
\mathcal{E}^{\prime }$ satisfying the condition $\bar{\rho}_{\mathcal{E}%
^{\prime }}=\bar{\rho}_{\mathcal{E}}$.

Now define the POVM dual to ensemble (\ref{GE}), (\ref{GE2}):
\begin{equation}
M^{\prime }(d^{2s}z)=\bar{\rho}_{\mathcal{E}}^{-1/2}\rho _{\beta ,z}\bar{\rho%
}_{\mathcal{E}}^{-1/2}\pi _{\gamma }(d^{2s}z)=D(\tilde{K}z)\rho _{\tilde{%
\beta}}D(\tilde{K}z)^{\ast }\frac{\left\vert \det \tilde{K}\right\vert
d^{2s}z}{\left( 2\pi \right) ^{s}},  \label{key}
\end{equation}%
where $\tilde{K}$ is a nondegenerate matrix (given explicitly by (\ref%
{ktilde})). The second equality follows with the help of results in ~\cite%
{lami}, Sec. 3.2 (see also Appendix 1). In particular, for $z=0$ it amounts
to $\rho _{\tilde{\beta}}\sim \rho _{\tilde{\alpha}}^{-1/2}\rho _{\beta
}\rho _{\tilde{\alpha}}^{-1/2},$ or $\rho _{\tilde{\alpha}}^{1/2}\rho _{%
\tilde{\beta}}\rho _{\tilde{\alpha}}^{1/2}\sim \rho _{\beta }$ ($\sim $
means \textquotedblleft proportional\textquotedblright ). The correlation
matrix of the operator $\rho _{1}^{1/2}\rho _{2}\rho _{1}^{1/2}$ where $\rho
_{1},\rho _{2}$ are Gaussian
is given in ~\cite{lami}, eq. (3.27), see also Corollary \ref{cor4} in Appendix 1 . In our case ($\rho _{1}=\bar{\rho}_{%
\mathcal{E}}=\rho _{\tilde{\alpha}}$, $\rho _{2}=\rho _{\tilde{\beta}}$) it
reads 
\begin{equation}
\beta =\tilde{\alpha}-\tilde{\Upsilon}\tilde{\alpha}\left( \tilde{\beta}+%
\tilde{\alpha}\right) ^{-1}\tilde{\alpha}\tilde{\Upsilon}^{t}.  \label{121}
\end{equation}%
Reversing (\ref{121}) and using $\tilde{\alpha}-\beta =\gamma ,$ we get
\begin{equation}
\tilde{\beta}=\tilde{\Upsilon}\tilde{\alpha}\gamma ^{-1}\tilde{\alpha}\tilde{%
\Upsilon}^{t}-\tilde{\alpha}.  \label{rev}
\end{equation}%
By noticing that $\tilde{\Upsilon}\tilde{\alpha}=\tilde{\alpha}\tilde{%
\Upsilon}^{t},$ see ~\cite{lami}, Eq. (3.22)-(3.25), we arrive at (\ref%
{tildeb}). Then by (\ref{inf}) and by lemma \ref{l1} above%
\begin{equation*}
A(\mathcal{E})=\sup_{M}I(\mathcal{E},M)=\max_{\mathcal{E}^{\prime }:\bar{\rho%
}_{\mathcal{E}^{\prime }}=\rho _{\tilde{\alpha}}}I(\mathcal{E}^{\prime
},M^{\prime })
\end{equation*}%
\begin{equation}
=\frac{1}{2}\log \mathrm{\det }\left( \tilde{\alpha}+\tilde{\beta}\right)
\left( \tilde{\beta}+\frac{1}{2}\Delta J_{\tilde{\beta}}\right) ^{-1},
\label{ag}
\end{equation}%
provided the condition (\ref{cond}) is fulfilled.

The statement concerning the optimal observable is obtained from the
corresponding statement of lemma \ref{l1} replacing $\alpha ,\beta $ by $%
\tilde{\alpha},\tilde{\beta}.$ Here the optimal ensemble consists of $J_{%
\tilde{\beta}}-$coherent states $D(z)\rho _{\frac{1}{2}\Delta J_{\tilde{\beta%
}}}D(z)^{\ast }$, and it is dual to the observable \ of the form (\ref{oo})
with some $K$ and $\rho _{\beta_{\ast }}\sim\rho _{\tilde{\alpha}%
}^{-1/2}\rho _{\frac{1}{2}\Delta J_{\tilde{\beta}}}\rho _{\tilde{\alpha}%
}^{-1/2}.$ By using (\ref{rev}) with $\gamma $ replaced by $\tilde{\alpha}-%
\frac{1}{2}\Delta J_{\tilde{\beta}}$ we obtain (\ref{opt}). $\square $

It is interesting to compare the quantity (\ref{ag}) with the lower bound
obtained by taking the heterodyne observable (\ref{vacM}). %
%
According to (\ref{plus}), the probability density of outcomes of this
observable for the Gaussian input state $\rho _{\tilde{\alpha}}$ is centered
Gaussian with the covariance matrix $\tilde{\alpha}+\frac{1}{2}\Delta
J_{\beta }=\gamma +\beta +\frac{1}{2}\Delta J_{\beta }=\alpha +\beta ,$
where at the last step we used (\ref{gamma}).

Computation using (\ref{hps}) and (\ref{mine}) gives the Shannon information%
\begin{eqnarray}
I(\mathcal{E},M) &=&h(p_{\rho _{\tilde{\alpha}}})-h(p_{\rho _{\frac{1}{2}%
\Delta J_{\beta }}})  \label{LE} \\
&=&\frac{1}{2}\log \mathrm{\det }\left( \alpha +\beta \right) \left( \beta +%
\frac{1}{2}\Delta J_{\beta }\right) ^{-1}.  \notag
\end{eqnarray}%
for the ensemble $\mathcal{E}$ and observable $M$ defined by (\ref{vacM})\
thus giving a lower bound for the accessible information $A(\mathcal{E}).$

We thus have the inequality between (\ref{aim}) and the lower bound (\ref{LE}%
) 
\begin{equation}
\frac{1}{2}\log \mathrm{\det }\left( \alpha +\beta \right) \left( \beta +%
\frac{1}{2}\Delta J_{\beta }\right) ^{-1}\leq \frac{1}{2}\log \mathrm{\det }%
\left( \tilde{\alpha}+\tilde{\beta}\right) \left( \tilde{\beta}+\frac{1}{2}%
\Delta J_{\tilde{\beta}}\right) ^{-1},  \label{que}
\end{equation}%
which becomes equality in the gauge-invariant case.

\begin{figure}[t]
\center{\includegraphics[width=0.5\textwidth]{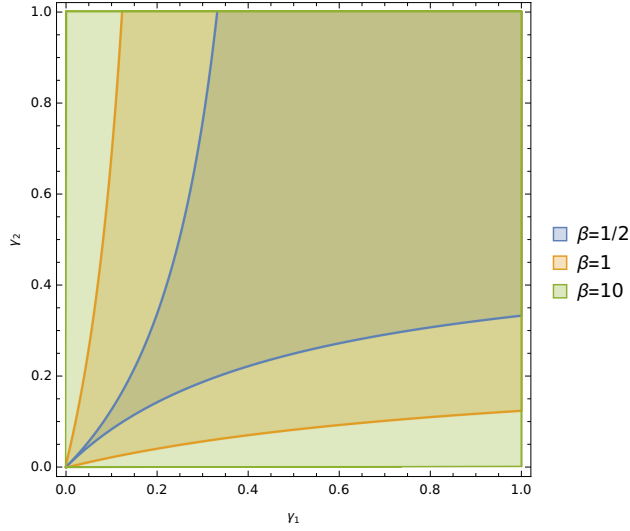}}
\caption{(color online) The ``threshold condition'' domain for $\protect\beta%
=1/2, 1, 10$.}
\label{plotAll}
\end{figure}


\section{One mode}

\label{s3}

We start with the case of lemma \ref{l1}. Let the measurement noise
covariance matrix be
\begin{equation}
\beta =\left[
\begin{array}{cc}
\beta _{1} & 0 \\
0 & \beta _{2}%
\end{array}%
\right] ;\quad \beta _{1}\beta _{2}\geq \frac{1}{4}.  \label{beta1}
\end{equation}%
The corresponding complex structure is
\begin{equation*}
J_{\beta }=\left[
\begin{array}{cc}
0 & -\sqrt{\beta _{2}/\beta _{1}} \\
\sqrt{\beta _{1}/\beta _{2}} & 0%
\end{array}%
\right] ,
\end{equation*}%
Notice, that when $\beta _{1}=\beta _{2},$ we are in the gauge-invariant
case with the standard complex structure
\begin{equation*}
J=\left[
\begin{array}{cc}
0 & -1 \\
1 & 0%
\end{array}%
\right] .
\end{equation*}

The covariance matrix of the squeezed vacuum is
\begin{equation*}
\frac{1}{2}\Delta J_{\beta }=\frac{1}{2}\left[
\begin{array}{cc}
\sqrt{\beta _{1}/\beta _{2}} & 0 \\
0 & \sqrt{\beta _{2}/\beta _{1}}%
\end{array}%
\right] .
\end{equation*}%
and
\begin{equation*}
\beta +\frac{1}{2}\Delta J_{\beta }=\left[
\begin{array}{cc}
\beta _{1}+\frac{1}{2}\sqrt{\beta _{1}/\beta _{2}} & 0 \\
0 & \beta _{2}+\frac{1}{2}\sqrt{\beta _{2}/\beta _{1}}%
\end{array}%
\right] ,
\end{equation*}%
so that $\det \left( \beta +\frac{1}{2}\Delta J_{\beta }\right) =\left(
\sqrt{\beta _{1}\beta _{2}}+1/2\right) ^{2},$ hence the second term in the
information quantity (\ref{CXA}) is $-\log \left( \sqrt{\beta _{1}\beta _{2}}%
+1/2\right) .$

Let us restrict to the diagonal input covariance matrices
\begin{equation*}
\alpha =\left[
\begin{array}{cc}
\alpha _{1} & 0 \\
0 & \alpha _{2}%
\end{array}%
\right] ,\quad \quad \alpha _{1}\alpha _{2}\geq \frac{1}{4}.
\end{equation*}%
Then the condition (\ref{thr}) amounts to $\alpha _{1}\geq \frac{1}{2}\sqrt{%
\beta _{1}/\beta _{2}},\quad \alpha _{2}\geq \frac{1}{2}\sqrt{\beta
_{2}/\beta _{1}},$ or%
\begin{equation}
\frac{1}{4\alpha _{2}^{2}}\leq \frac{\beta _{1}}{\beta _{2}}\leq 4\alpha
_{1}^{2}.  \label{thr0}
\end{equation}%
The matrix
\begin{equation*}
\alpha +\beta =\left[
\begin{array}{cc}
\beta _{1}+\alpha _{1} & 0 \\
0 & \beta _{2}+\alpha _{2}%
\end{array}%
\right]
\end{equation*}%
has the determinant $\left( \beta _{1}+\alpha _{1}\right) \left( \beta
_{2}+\alpha _{2}\right) ,$ so that the information quantity (\ref{CXA}) (and
the lower bound in (\ref{LE})) is%
\begin{equation}
\max_{\mathcal{E}:\bar{\rho}_{\mathcal{E}}=\rho_{\alpha }}I(%
\mathcal{E},\tilde{M})=\frac{1}{2}\log \frac{\left( \beta _{1}+\alpha
_{1}\right) \left( \beta _{2}+\alpha _{2}\right) }{\left( \sqrt{\beta
_{1}\beta _{2}}+1/2\right) ^{2}}.  \label{lb}
\end{equation}

Let us now turn to the theorem \ref{t1} for a Gaussian ensemble $\mathcal{E}%
=\left\{ \pi _{\gamma }(d^{2}z),\rho _{\beta ,z}\right\} $ with the diagonal
covariance matrices
\begin{equation*}
\gamma =\left[
\begin{array}{cc}
\gamma _{1} & 0 \\
0 & \gamma _{2}%
\end{array}%
\right] \geq 0
\end{equation*}%
and $\beta $ of the form (\ref{beta1}). By (\ref{tildea})
\begin{equation}
\tilde{\alpha}_{1}=\beta _{1}+\gamma _{1},\quad \tilde{\alpha}_{2}=\beta
_{2}+\gamma _{2}.  \label{tildea1}
\end{equation}%
Let us find the matrix
\begin{equation*}
\tilde{\beta}=\left[
\begin{array}{cc}
\tilde{\beta}_{1} & 0 \\
0 & \tilde{\beta}_{2}%
\end{array}%
\right] .
\end{equation*}%
According to (\ref{tildeb}) we have
\begin{eqnarray}
\tilde{\beta}_{1} &=&\frac{\left( \beta _{1}+\gamma _{1}\right) }{\gamma _{1}%
}\left[ \beta _{1}-\frac{1}{4\left( \beta _{2}+\gamma _{2}\right) }\right] ,
\notag \\
\tilde{\beta}_{2} &=&\frac{\left( \beta _{2}+\gamma _{2}\right) }{\gamma _{2}%
}\left[ \beta _{2}-\frac{1}{4\left( \beta _{1}+\gamma _{1}\right) }\right] .
\label{tildeb1}
\end{eqnarray}%
Note that $\beta _{1}\beta _{2}\geq 1/4$ implies $\tilde{\beta}_{1}\tilde{%
\beta}_{2}\geq 1/4$ 
as required for the covariance matrix of a Gaussian state. By (\ref{thr0})
the condition (\ref{cond}) amounts to%
\begin{equation}
\frac{1}{4\tilde{\alpha}_{2}^{2}}\leq \frac{\tilde{\beta}_{1}}{\tilde{\beta}%
_{2}}\leq 4\tilde{\alpha}_{1}^{2}.  \label{thr1}
\end{equation}%
The accessible information (\ref{aim}) is 
\begin{equation}
A(\mathcal{E})=\frac{1}{2}\log \frac{\left( \tilde{\alpha}_{1}+\tilde{\beta}%
_{1}\right) \left( \tilde{\alpha}_{2}+\tilde{\beta}_{2}\right) }{\left(
\sqrt{\tilde{\beta}_{1}\tilde{\beta}_{2}}+1/2\right) ^{2}}.  \label{acc1}
\end{equation}%

To obtain the expressions in terms of the ensemble parameters $\gamma ,\beta
,$ one must substitute the relations (\ref{tildea1}), (\ref{tildeb1}) into (%
\ref{thr1}), (\ref{acc1}). After some calculations which are done in the
Appendix 2 we obtain the threshold condition%
\begin{equation}
\frac{1}{4\left( \beta _{2}+\gamma _{2}\right) \beta _{2}}\leq \frac{\gamma
_{1}}{\gamma _{2}}\leq 4\left( \beta _{1}+\gamma _{1}\right) \beta _{1}
\label{thr2}
\end{equation}%
and the accessible information%
\begin{equation}
A(\mathcal{E})=\log \frac{\left[ \left( \beta _{1}+\gamma _{1}\right) \left(
\beta _{2}+\gamma _{2}\right) -\frac{1}{4}\right] }{\sqrt{\left[ \left(
\beta _{1}+\gamma _{1}\right) \beta _{2}-\frac{1}{4}\right] \left[ \left(
\beta _{2}+\gamma _{2}\right) \beta _{1}-\frac{1}{4}\right] }+\frac{\sqrt{%
\gamma _{1}\gamma _{2}}}{2}}.  \label{acc2}
\end{equation}

Computation of the parameters (\ref{opt}) of the optimal Gaussian observable
(\ref{oo}) gives%
\begin{eqnarray*}
\beta _{\ast 1}&=&\frac{1}{2}\sqrt{\frac{\tilde{\beta}_{1}}{\tilde{\beta}_{2}%
}}\frac{\tilde{\alpha}_{1}}{\tilde{\alpha}_{2}}\left( \frac{\tilde{\alpha}%
_{2}-\frac{1}{2}\sqrt{\tilde{\beta}_{2}/\tilde{\beta}_{1}}}{\tilde{\alpha}%
_{1}-\frac{1}{2}\sqrt{\tilde{\beta}_{1}/\tilde{\beta}_{2}}}\right) , \\
\beta _{\ast 2}&=&\frac{1}{2}\sqrt{\frac{\tilde{\beta}_{2}}{\tilde{\beta}_{1}%
}}\frac{\tilde{\alpha}_{2}}{\tilde{\alpha}_{1}}\left( \frac{\tilde{\alpha}%
_{1}-\frac{1}{2}\sqrt{\tilde{\beta}_{1}/\tilde{\beta}_{2}}}{\tilde{\alpha}%
_{2}-\frac{1}{2}\sqrt{\tilde{\beta}_{2}/\tilde{\beta}_{1}}}\right) .
\end{eqnarray*}%
Notice that $\beta _{\ast 1}\beta _{\ast 2}=1/4$ as it must be for a
squeezed vacuum.

To simplify visualization of the condition (\ref{thr1}) we can assume
without loss of generality (via a symplectic coordinate transformation) that
$\beta _{1}=\beta _{2}=\beta \geq 1/2$. Then the sets of solutions $\left(
\gamma _{1},\gamma _{2}\right) $ of the system (\ref{thr1}) for $\beta=1/2,
1, 10,$ are shown on Fig. \ref{plotAll}~ ~\cite{note}.

The inequality (\ref{que}) becomes 
\begin{equation*}
\frac{1}{2}\log \frac{\left( \beta +\gamma _{1}+1/2\right) \left( \beta
+\gamma _{2}+1/2\right)}{ \left( \beta +1/2\right) ^{2}}
\end{equation*}%
\begin{equation*}
\leq \frac{1}{2}\log \frac{\left( \beta +\gamma _{1}+\tilde{\beta}%
_{1}\right) \left( \beta +\gamma _{2}+\tilde{\beta}_{2}\right)}{ \left(
\sqrt{\tilde{\beta}_{1}\tilde{\beta}_{2}}+1/2\right) ^{2}},
\end{equation*}
which turns into equality iff $\gamma _{1}=\gamma _{2}$ (the gauge-invariant
case).

Examples of ensemble not satisfying the key condition (\ref{cond}) of
theorem \ref{t1} are obtained by taking the parameters $\beta \geq
1/2,\gamma _{1}\geq 0,\gamma _{2}\geq 0,$ not satisfying at least one of the
inequalities (\ref{thr2}) (outer domains of curved angles on Fig. \ref%
{plotAll}). \ A notable case is $\gamma _{1}>0,\gamma _{2}=0,$ which
corresponds to the ensemble with the Gaussian distribution $\pi _{\gamma
_{1}}(dx)$ concentrated on the horizontal axis $x,$ and the family of states
\begin{equation*}
\rho _{\beta ,x}=D(x,0)\rho _{\beta }D(x,0)^{\ast },
\end{equation*}%
where $D(x,0)=\exp (-ixp)$ are the position displacement operators and $\rho
_{\beta }$ is the gauge-invariant Gaussian (thermal) state. Theorem \ref{t1}
does not apply in this case while a natural conjecture is that the optimal
measurement for the accessible information of this ensemble is still
\textquotedblleft Gaussian\textquotedblright\ (namely, the sharp position
measurement, cf. sec. 5 of the paper~\cite{entropy}).

\begin{acknowledgments}
The work was supported by the grant of Russian
Science Foundation (project No 19-11-00086).
The author is grateful to Vsevolod Yashin for useful comments and the help
with graphics.
\end{acknowledgments}



\section*{Appendix 1}

In our notations the statement of Lemma 5 of the paper ~\cite{lami} reads
\begin{equation}
\mathrm{Tr}\,W(z_{1})\sqrt{\rho _{\alpha }}W(-z_{2})\sqrt{\rho _{\alpha }}%
=\exp \left( -\frac{1}{2}z_{2}^{t}\alpha z_{2}-\frac{1}{2}z_{1}^{t}\alpha
z_{1}+z_{2}^{t}\kappa z_{1}\right) ,  \label{sqsq}
\end{equation}%
where
\begin{equation}
\kappa =\sqrt{I_{2s}+\left( 2\alpha \Delta ^{-1}\right) ^{-2}}\,\alpha
=\alpha \sqrt{I_{2s}+\left( 2\Delta ^{-1}\alpha \right) ^{-2}}.  \label{kap}
\end{equation}%

\textit{Sketch of proof}. The quantum Fourier transform of $\sqrt{\rho
_{\alpha }}$ computed in ~\cite{tmf} is
\begin{eqnarray}
f(w) &=&\mathrm{Tr}\,\sqrt{\rho _{\alpha }}W(w)  \label{sqr} \\
&=&\sqrt[4]{\det \left( 2\hat{\alpha}\right) }\exp \left( -\frac{1}{2}w^{t}%
\hat{\alpha}w\right) ,  \notag
\end{eqnarray}%
where
\begin{equation*}
\hat{\alpha}=\alpha +\kappa =\alpha \left( I_{2s}+\sqrt{I_{2s}+\left(
2\Delta ^{-1}\alpha \right) ^{-2}}\right) .
\end{equation*}%
Hence 
\begin{equation*}
\mathrm{Tr}\,(W(z_{1})\sqrt{\rho _{\alpha }})\,W(w)=\mathrm{Tr}\,\sqrt{\rho
_{\alpha }}W(w)W(z_{1})=\exp \left( -\frac{i}{2}w^{t}\Delta z_{1}\right)
f(w+z_{1}).
\end{equation*}%
By using Parceval relation for the quantum Fourier transform ~\cite{aspekty}%
, we have
\begin{eqnarray*}
&&\mathrm{Tr}\,W(z_{1})\sqrt{\rho _{\alpha }}W(-z_{2})\sqrt{\rho _{\alpha }}
\\
&=&\mathrm{Tr}\,\left( W(z_{1})\sqrt{\rho _{\alpha }}\right) \left( \sqrt{%
\rho _{\alpha }}W(z_{2})\right) ^{\ast } \\
&=&\frac{1}{\left( 2\pi \right) ^{s}}\int \exp \left( -\frac{i}{2}%
w^{t}\Delta z_{1}\right) f(w+z_{1})\overline{\exp \left( \frac{i}{2}%
w^{t}\Delta z_{2}\right) f(w+z_{2})}d^{2s}w \\
&=&\frac{1}{\left( 2\pi \right) ^{s}}\int \exp \left[ -\frac{i}{2}%
w^{t}\Delta \left( z_{1}+z_{2}\right) \right] f(w+z_{1})\overline{f(w+z_{2})}%
d^{2s}w.
\end{eqnarray*}%
Substituting (\ref{sqr}), computing a Gaussian integral and
using the relation
\begin{equation*}
\hat{\alpha}-\frac{1}{4}\Delta \hat{\alpha}^{-1}\Delta =2\alpha
\end{equation*}%
from the paper ~\cite{scu} gives (\ref{sqsq}). $\square $


\begin{corollary}\label{cor4}
The following relation holds
\begin{equation}\label{cor}
\mathrm{Tr}\,(\sqrt{\rho _{\alpha }}\rho _{\beta ,z}\sqrt{\rho _{\alpha }}%
)W(z_{1})=c\,\exp \left( iz_{1}^{t}Kz-\frac{1}{2}z_{1}^{t}\alpha
_{121}z_{1}\right) ,
\end{equation}%
where
\begin{equation*}
c=\left( \det \left( \alpha +\beta \right) \right) ^{-1/2}\exp \left( -\frac{%
1}{2}z^{t}\left( \alpha +\beta \right) ^{-1}z\right) ,
\end{equation*}%
\begin{eqnarray*}
\alpha _{121} &=&\alpha -\kappa \left( \alpha +\beta \right) ^{-1}\kappa
,\quad  \\
K &=&\kappa \left( \alpha +\beta \right) ^{-1}.
\end{eqnarray*}
\end{corollary}

We mention in passing that the characteristic function of the product of
Gaussian density operators was obtained in ~\cite{hsh}.

\textit{Proof.} By the inversion formula for the quantum Fourier transform ~%
\cite{aspekty} 
\begin{equation*}
\rho _{\beta ,z}=\frac{1}{\left( 2\pi \right) ^{s}}\int \exp \left(
iz_{2}^{t}z-\frac{1}{2}z_{2}^{t}\beta z_{2}\right) W(-z_{2})d^{2s}z_{2}.
\end{equation*}%
Combining with (\ref{sqsq}), 
\begin{eqnarray*}
&&\mathrm{Tr}\,\sqrt{\rho _{\alpha }}\rho _{\beta ,z}\sqrt{\rho _{\alpha }}%
W(z_{1}) \\
=\frac{1}{\left( 2\pi \right) ^{s}}\int \exp \left( iz_{2}^{t}z-\frac{1}{2}%
z_{2}^{t}\beta z_{2}\right)  &&\exp \left( -\frac{1}{2}z_{2}^{t}\alpha z_{2}-%
\frac{1}{2}z_{1}^{t}\alpha z_{1}+z_{2}^{t}\kappa z_{1}\right) d^{2s}z_{2}.
\end{eqnarray*}%
Computation of the Gaussian integral results in (\ref{cor}). $\square $

Replacing in (\ref{cor}), (\ref{kap}) $\alpha ,\beta $ by $\tilde{\alpha},%
\tilde{\beta}$, we rederive (\ref{121}). Replacing additionally $z$ by $%
\tilde{K}z,$ where 
\begin{equation}
\tilde{K}=\left( \tilde{\alpha}+\tilde{\beta}\right) \tilde{\kappa}%
^{-1}=\left( \tilde{\alpha}+\tilde{\beta}\right) \tilde{\alpha}^{-1}\tilde{%
\Upsilon}^{-1}\,  \label{ktilde}
\end{equation}%
after some routine calculations using (\ref{rev}) we obtain
\begin{equation*}
\mathrm{Tr}\,(\sqrt{\rho _{\tilde{\alpha}}}\rho _{\tilde{\beta},\tilde{K}z}%
\sqrt{\rho _{\tilde{\alpha}}})W(z_{1})=\frac{1}{\left\vert \det \tilde{K}%
\,\right\vert \sqrt{\det \gamma }}\,\exp \left( -\frac{1}{2}z^{t}\gamma
^{-1}z\right) \mathrm{Tr}\,\rho _{\beta ,z}W(z_{1}),
\end{equation*}%
implying (\ref{key}).

\section*{Appendix 2}

\textit{Proof of (\ref{thr2}), (\ref{acc2}). }The second inequality in (\ref%
{thr1}) is the same as $\tilde{\beta}_{1}\leq 4\tilde{\alpha}_{1}^{2}\tilde{%
\beta}_{2}.$ By using (\ref{tildea1}), (\ref{tildeb1}), we obtain%
\begin{equation*}
\tilde{\alpha}_{1}-\frac{1}{4\tilde{\alpha}_{2}}-\gamma _{1}\leq 4\tilde{%
\alpha}_{1}\tilde{\alpha}_{2}\frac{\gamma _{1}}{\gamma _{2}}\left( \tilde{%
\alpha}_{2}-\frac{1}{4\tilde{\alpha}_{1}}-\gamma _{2}\right) ,
\end{equation*}%
or, introducing $D=$ $\tilde{\alpha}_{1}\tilde{\alpha}_{2}-\frac{1}{4},$%
\begin{equation*}
\frac{D}{\tilde{\alpha}_{2}}-\gamma _{1}\leq 4\tilde{\alpha}_{2}\frac{\gamma
_{1}}{\gamma _{2}}D-4\tilde{\alpha}_{1}\tilde{\alpha}_{2}\gamma _{1}.
\end{equation*}%
Rearranging and dividing by $D>0,$%
\begin{equation*}
\frac{1}{\tilde{\alpha}_{2}}\leq 4\tilde{\alpha}_{2}\frac{\gamma _{1}}{%
\gamma _{2}}-4\gamma _{1}=4\gamma _{1}\left( \frac{\tilde{\alpha}_{2}}{%
\gamma _{2}}-1\right) =4\frac{\gamma _{1}}{\gamma _{2}}\left( \tilde{\alpha}%
_{2}-\gamma _{2}\right) ,
\end{equation*}%
which is the same as%
\begin{equation*}
\frac{1}{4\tilde{\alpha}_{2}\left( \tilde{\alpha}_{2}-\gamma _{2}\right) }%
\leq \frac{\gamma _{1}}{\gamma _{2}},
\end{equation*}%
equivalent to the first inequality in (\ref{thr2}) by (\ref{tildea1}).

Again by using (\ref{tildea1}), (\ref{tildeb1}), we obtain%
\begin{equation*}
\tilde{\alpha}_{1}+\tilde{\beta}_{1}=\frac{\tilde{\alpha}_{1}}{\gamma _{1}%
\tilde{\alpha}_{2}}\left( \tilde{\alpha}_{1}\tilde{\alpha}_{2}-\frac{1}{4}%
\right) ,\quad \tilde{\alpha}_{2}+\tilde{\beta}_{2}=\frac{\tilde{\alpha}_{2}%
}{\gamma _{2}\tilde{\alpha}_{1}}\left( \tilde{\alpha}_{1}\tilde{\alpha}_{2}-%
\frac{1}{4}\right) ,
\end{equation*}%
hence%
\begin{equation*}
\left( \tilde{\alpha}_{1}+\tilde{\beta}_{1}\right) \left( \tilde{\alpha}_{2}+%
\tilde{\beta}_{2}\right) =\frac{1}{\gamma _{1}\gamma _{2}}\left( \tilde{%
\alpha}_{1}\tilde{\alpha}_{2}-\frac{1}{4}\right) ^{2}.
\end{equation*}%
Moreover,%
\begin{equation*}
\tilde{\beta}_{1}\tilde{\beta}_{2}=\frac{1}{\gamma _{1}\gamma _{2}}\left(
\tilde{\alpha}_{1}\beta _{2}-\frac{1}{4}\right) \left( \tilde{\alpha}%
_{2}\beta _{1}-\frac{1}{4}\right) .
\end{equation*}%
Substituting into (\ref{acc1}) we get (\ref{acc2}).

\end{document}